	\documentclass[12pt]{article}
	\usepackage{geometry}                
	\usepackage{amssymb}
	\usepackage{amsmath}
	\usepackage{amsbsy}
	\usepackage{fourier}

	\usepackage[english]{babel}
	\usepackage{authblk}	
	\usepackage{bm}
	\usepackage{bbold}
	\usepackage{mathrsfs}
	\usepackage{hyphsubst}
	\usepackage{sectsty}
	\usepackage{lettrine}
	\usepackage[margin=45pt]{caption}

	\usepackage{cite}

	\usepackage{graphicx}
	\usepackage[caption=false]{subfig}
	\usepackage{float}

	\usepackage{nextpage}
	\usepackage[comma,numbers,sort&compress]{natbib}

\title{		Squeezing: from numerical to conceptual problems				}
\author{	B. Mielnik\footnote{bogdan@fis.cinvestav.mx}~  and J. Fuentes\footnote{jfuentes@fis.cinvestav.mx}									}
\affil{\small	Department of Physics, CINVESTAV-IPN, \\  A.P. 14-740, Mexico City 07000. Mexico.    }
\date{}

\begin{document}


\newpage \setcounter{page}{1} \maketitle



\begin{abstract}{\footnotesize

\noindent In the studies of the squeezing it is customary to focus more attention on the particular squeezed states and their evolution than on the dynamical operations that could squeeze simultaneously some wider families of quantum states, independently of their initial shape. We look for new steps in this direction, carried out by softly acting external fields which might produce the squeezing of the canonical observables $q,p$ of charged particles. The works on these problems collect so many valuable results that the question is whether something more is indeed something  more in our knowledge. Yet we decided to present some exactly solvable cases of the problem which appear in the symmetric evolution intervals permitting to find explicitly the time dependence of the external fields needed to generate the required evolution operators. Curiously, our results are interrelated with a simple anti-commuting algebra of Toeplitz which describes the problem more easily than the frequently used Ermakov--Milne invariants. Some trending topics, as well as some fundamental problems in quantum theory, might also be involved.}
\end{abstract}


{\small \noindent \hspace{.8cm} PACS numbers: 03.65.Fd, 03.67.Ta}

{\small \noindent \hspace{.8cm} Keywords: quantum control, dynamical operations, squeezing 
}

\newpage
\section{Introduction}
\label{sec:intro}

Traditionally, in some basic control problems, two pictures of quantum theory are competing: that of Schr\"odinger (state evolution) and of Heisenberg (the evolution of observables). The advantages of Heisenberg's representation were strongly defended in a polemic article of Dirac  \cite{dirac65}. Even if one does not share his criticism, some of its practical aspects revive in the quantum control problems. It concerns specially the linear transformations of the canonical variables $q,p$ including the squeezing, as described by Yuen \cite{yuen76}. One of its applications shows the possibility of an exact and fast measurement of the free particle position by using the twisted canonical observables \cite{yuen83}. By looking carefully, his squeezing techniques were the first glimpse into the future quantum tomography \cite{mancini96,manciniar,asorey11} describing the quantum states in terms of tomographic images defined via the Radon transforms of the canonical variables \cite{radon17,dunajski}, a domain recently represented by a growing avalanche of ideas  and contributions \cite{Seth}.  Some consequences of this view for the basic Quantum Mechanical ideas, though not immediate, are not negligible.

In the formalism of quantum mechanics every self-adjoint operator in the Hilbert space represents a measuring device which performs an instantaneous check, reducing the wave packet to one of the eigenstates. Yet, the laboratory practice suggests a more complex image in which the act of measurement is prepared by a previous (auxiliary) evolution and the fundamental (indeterministic) choice occurs at the very end.

Below, we shall be specially interested in the auxiliary operations of amplification or squeezing as a chance to apply the ideas of the demolition free measurements of Thorne et al. \cite{nondemolishing78,nondemolishing80}. Of course, not all techniques available awake an absolute   confidence, including the decoherence \cite{bohm66}, the instability \cite{haroche98}, and the 'delayed choice' \cite{wheeler84}. Recently, even the entangled states and their radiation effects are under some  critical attention \cite{Lloyd12,ma,dolev,price}. 

For these reasons, we shall deal only with the soft control techniques, without sudden jumps and radiative  pollution. Our problem though  elementary, is still not completely trivial. In fact, in the known theory it is not even granted that each well defined unitary operator can be dynamically achieved (or at least approximated) by realistic motion generators. We therefore design a simple class of time dependent Hamiltonians, trying to find out whether they could generate the true squeezing effects of massive particles. We consider the variable external fields as the only credible source of such phenomenon. So, we skip all formal results obtained for time dependent masses, material constants, etc. Yet, the number of known results still impressing  (cf. the encyclopedic report by V. Dodonov \cite{dodonov02}). We also decided  to avoid the tentative dangers of QFT  \cite{wheeler84,Lloyd12}. Henceforth we try to keep our prescriptions on purely quantum mechanical level, restricting our operations to slow (adiabatic) processes with minimal radiative pollution. 

Our report is organized as follows. Section 2 presents the 'trajectory method' which permits to deduce the quantum evolution generated by the quadratic Hamiltonians from the classical motion trajectories. In sections 3 and 4 we present and classify the analogues of optical operations for massive particles. Section 5 is dedicated to an extra material, on the generally unnoticed squeezing effects in the experimental conditions traditional for Paul's traps. Section 6 discusses our principal idea of the squeezing by the distorted Fourier transformations caused by sharp pulses of the oscillator potentials. Even if our solutions do not yet provide the exact laboratory prescriptions, they indicate that the squeezing effects can be experimentally approximated. This is thanks to the exact solutions described in sections 7--9,  where we report an elementary case of the Toeplitz algebra which permits to design the exact, soft equivalents of the desired dynamical effects. On section 10 we analyze their possible limitations, and finally, section 11 reports some fundamental hopes but also difficulties.

For convenience, our mathematical calculations are carried in dimensionless variables but the results are then translated into the physical units.


\section{The Classical--Quantum Duality}
\label{sec:duality}

\noindent The simplest  phenomena of {\em squeezing} can occur in the evolution  of canonical variables $q,p$ for the non-relativistic time dependent, quadratic Hamiltonians in 1D with variable elastic forces in either classical or quantum theory:
\begin{equation}
\label{eq:1}
H(\tau) = \frac{p^2}{2}+\beta(\tau)\frac{q^2}{2}
\end{equation}
where,  $q, p$ are the dimensionless canonical position and momentum,  $\tau$  is a dimensionless time, and we adopt units in which the mass $m = 1$; in quantum case also  $\hbar = 1$, $[q,p] = i$. Our question is, whether the evolution generated by the Hamiltonian \eqref{eq:1} can at some moment produce a unitary operator $U$ transforming $q\rightarrow \lambda q,\ p\rightarrow \frac{1}{\lambda} p$ with $0\neq\lambda\in\mathbb{R}$, {\em i.e.} squeezing $q$ and expanding $p$ or {\it vice versa}? A more general question is, whether for any pair of quantum observables $a=a^\dagger$ and  $b=b^\dagger$, commuting to a number $[a,b]=i\alpha$ $(\alpha\in\mathbb{R})$ an evolution operator can  transform $ a\rightarrow \lambda a,\  b\rightarrow \frac{1}{\lambda} b$, {\em i.e.} expanding $a$ at the cost of $b$ 
or inversely? 
\bigskip 

The behavior of non-relativistic particles in 1D in variable oscillatory fields \eqref{eq:1} was studied with the aim to describe the particle motion in Paul's traps   
(cf. Paul and collaborators \cite{paul90}), then in ample contributions of Glauber \cite{glauber92} and others. 
Some results about the  {\em squeezing} caused by variable electromagnetic fields were studied by Baseia et al \cite{baseia92,baseia93}.   Can one still achieve something more? As it seems, in some circumstances, instead of 
using the Wronskian to compare the independent solutions of \eqref{eq:1} an easier method might be to use the {\em classical--quantum duality} permitting to deduce  the quantum evolution operators 
from the classical motion. The method could not work  for the general quantum motion, but it does 
for all quadratic Hamiltonians including  \eqref{eq:1}.   
 
Notice that for nonsingular, bounded $\beta(\tau)$, the evolution equations generated by \eqref{eq:1} in both classical and quantum cases imply exactly the same linear equations for either classical or quantum canonical variables: $dq/d\tau= p(\tau)$, $dp/d\tau =-\beta(\tau)q(\tau)$, leading in any time interval $[\tau_0, \tau]$ to the identical  transformation of either classical or quantum canonical  pair, expressed by the same family of $2\times2$ symplectic {\em evolution matrices}  $u(\tau,\tau_0)$:
\begin{equation}
\label{eq:2}
\begin{pmatrix} 
q(\tau) \\ p(\tau)
\end{pmatrix} = u(\tau,\tau_0) 
\begin{pmatrix} q(\tau_0)\\p(\tau_0)\end{pmatrix}; \quad u(\tau_0,\tau_0) = 1,
\end{equation}
determined  by the matrix equations
\begin{equation}
\label{eq:3}
\frac{\text{d}}{\text{d}\tau} u(\tau,\tau_0) = \Lambda(\tau)u(\tau,\tau_0); \quad \Lambda(\tau) = \begin{pmatrix}0&&1\\-\beta(\tau)&&0\end{pmatrix}.
\end{equation}
The reciprocity between the classical and quantum pictures does not end up here.
It turns out that, in absence of spin, each unitary evolution operator $U(\tau,\tau_0)$ in $L^2(\mathbb{R})$ generated by the time dependent, quadratic Hamiltonian \eqref{eq:1} is determined, up to a phase factor, by the canonical transformation that it induces. This is the consequence of the following simple  
lemma  \cite{reed75,mielnik77,mielnik11,mielnik13}: 
\bigskip

{\bf Lemma 1}. The family of the unitary operators  $U(\tau,\tau_0)$
describing the evolution generated by the quadratic Hamiltonians \eqref{eq:1} is determined with accuracy to the $c$-number phase factors by the corresponding matrices  $u(\tau,\tau_0)$.   
 (And therefore, also by the corresponding classical trajectories.) 

 {\bf Proof.} Indeed, it is enough to notice that if two unitary operators $U_1$ and $U_2$ produce the same transformation of the canonical variables {\em i.e.} $U_1^{\dagger}qU_1 = U_2^{\dagger}qU_2$ and $U_1^{\dagger}pU_1 = U_2^{\dagger}pU_2$, then $U_1U_2^{\dagger}$ commutes with both $q$ and $p$. Hence, it commutes also with any function of $q$ and $p$. Since in $L^2(\mathbb{R})$ the functions of $q$ and $p$ generate an irreducible algebra, then $U_1U_2^\dagger$ must be a c-number and since it is unitary, it can be only a phase factor, $U_1U_2^\dagger =e^{i\varphi} \Rightarrow U_1 = e^{i\varphi}U_2$ where $\varphi \in \mathbb{R}$.

Now, any two unitary operators which differ only by a c-number phase, even if acting differently on the {\em state vectors}, generate the same transformation of {\em quantum states}, so we shall call them {\em equivalent}, $U_1\equiv U_2$. It follows immediately that the trajectories of the classical motion problem with the quadratic $H(\tau)$ determine completely the evolution of quantum pure or mixed states $\rho = \rho^\dagger  \geq 0$, $\text{Tr}~\rho = 1$ and, modulo equivalence, the entire unitary history,  which we denote for simplicity as $U(\tau)= U(\tau,\tau_0)$. So, in a sense, our description is complementary to the intriguing  trends of {\em phase geometry} \cite{berry84,anandan90,fernandez94,brody03,wolf04,carlini06,berry09,muga11,wolf12}. 
It does not describe the adiabatic or geometric phases, but it determines the alternative aspects of 
the quantum states like the motion of the canters, the packets shapes and all details of the statistical 
interpretation, including the  quantum uncertainties of all orders. 


\section{Elementary Models}
\label{sec:models}

Some traditional models illustrate  the above `duality doctrine'. Two of them seem of special interest.
\bigskip

(1) The evolution of charged particles in the hyperbolically shaped ion traps \cite{paul90}. The Paul's potentials $\Phi({\bf x},t)$ in the trap interior generated by the voltage $\Phi(t)$ on the surfaces are either $\Phi=\frac{e\Phi(t)}{r_0^2}\left(\frac{x^2}{2}+\frac{y^2}{2}-z^2\right)$ or $\Phi=\frac{e\Phi(t)}{r_0^2}\left(\frac{x^2}{2}-\frac{y^2}{2}\right)$. The problem then splits into the partial Hamiltonians of the type: $H(t)=\frac{p^2}{2}+\frac{e\Phi(t)}{r_0^2}\frac{q^2}{2}$, where $q,p$ represent just one of independent pairs of
 canonical observables. Now, by introducing the new dimensionless time variable $\tau = \frac{t}{T}$, where $T$ stands for an arbitrarily chosen {\em time scale}, each $1D$ Hamiltonian is reduced  to a particular case of \eqref{eq:1}:
\begin{equation}
\label{eq:4}
\tilde{H}(\tau) = H(t)T = \frac{\tilde{p}^2}{2}+\beta(\tau)\frac{\tilde{q}^2}{2}, \quad \beta(\tau) = \frac{e\Phi(t)T^2}{r_0^2 m}, 
\end{equation}
where $\beta(\tau)$ is dimensionless and the new  canonical variables $\tilde{q} =  \sqrt{\frac{m}{T}}q$ and $\tilde{p} = \sqrt{\frac{T}{m}} p$ are then expressed in the same units (square roots of the action), leading to the dimensionless evolution matrices $u(\tau,\tau_0)$ identical for the classical and quantum dynamics. So, {\em without even knowing} about the existence of quantum mechanics, the dimensionless quantities can be now constructed: 
\begin{equation}
\label{eq:5}
q_d=\frac{\tilde{q}}{\sqrt{\hbar}} = q\sqrt{\frac{m}{\hbar T}}, \quad p_d = \frac{\tilde{p}}{\sqrt{\hbar}} =p\sqrt{\frac{T}{\hbar m}}
\end{equation}
and $H_d(t) = \frac{H(t)T}{\hbar}$, where $\hbar$ is an arbitrarily chosen action unit ({\em cf.}~\cite{delgado98}). By knowing already about the quantum background of the theory, an obvious (though not obligatory) option is to choose $\hbar$ as the Planck constant (though the other constants  proportional to $\hbar$ of Planck are neither excluded\footnote{Indeed, for the quadratic Hamiltonians all results of {\em Mr. Tompkins in wonderland} by George Gamov, can be deduced just by rescaling time, canonical variables and the external fields.}). By dropping the unnecessary indexes, one ends up with the evolution problem \eqref{eq:1}, with an arbitrary time dependent $\beta(\tau)$ (not necessarily coinciding with  that of Paul\footnote{Note that the period of the external oscillating field  may, but needs not to be used as the 'reference time' $T$ to define the dimensionless variable $\tau =\frac{t}{T}$. In his known paper \cite{paul90} W. Paul assumes the voltage on the trap walls $\Phi(t)=\Phi_0+\Phi_1\cos\omega t$, choosing $\tau= \frac{\omega t}{2}$. Here, we assume simply $\tau = \omega t$ and use the Mathieu equation together with the stability  diagram of Strut in Bender-Orszag form \cite{bender78} to compare easily with \cite{mielnik10}. More general $\tau$-dependencies will be also considered.}).
\bigskip 

(2) The similar dynamical law applies to charged particles moving in a time dependent magnetic field, given (in the first step of Einstein--Infeld--Hoffmann (EIF) approximation \cite{infeld60}) by ${\bf B}(t ) = {\bf n} B(t)$, where ${\bf n}$ is a constant unit vector (defining the central $z-$axis of the cylindrical solenoid; see also relativistic equivalent in our section X). Since ${\bf B}(t)$ has the vector potential ${\bf A}(x,t) = \frac {1}{ 2} {\bf B}(t)\times{\bf x}$, the Hamiltonian describing the non-relativistic motion of the charge $e$ in the field ${\bf B}(t)$ is expressed  (in Gaussian units) by
\begin{equation}
\label{eq:6}
H(t) = \frac{1}{2m}\left({\bf p}-\frac{e}{c}{\bf A}\right)^2 =  \frac{1}{2m}\left[{\bf p}^2+\left(\frac{eB(t)}{2c}\right)^2(x^2+y^2)\right]-\frac{eB(t)}{2mc}M_z.
\end{equation}
After separating the free motion along the $z$-axis and the easily integrable rotations caused by $M_z = xp_y-yp_x$ (both commuting with $H(t)$), the motion of the non-relativistic charged, spinless particle on $2D$ plane $\mathcal{P}_\perp$ perpendicular to ${\bf n}$, obeys the simplified Hamiltonian
\begin{equation}
\label{eq:7}
H(t) = \frac{1}{2m}\left[{\bf p}^2+\left(\frac{eB(t)}{2c}\right)^2{\bf x}^2\right],
\end{equation}
where ${\bf p}$ and ${\bf x}$ are the pairs of canonical momenta and positions on $\mathcal{P}_\perp$. This, after using the dimensionless variables $\tau = \frac{t}{T}$, with $x,p_x$ and $y,p_y$ replacing $q,p$ in \eqref{eq:5}, leads again to a pair of motions of type \eqref{eq:1}, with the dimensionless $\tau, q , p$ and
\begin{equation}
\label{eq:8}
\beta(\tau)=\kappa^2(\tau)=\left(\frac{eTB(T\tau)}{2mc}\right)^2.
\end{equation}
In case, if $B (t )$ oscillates periodically with frequency $\omega$ the natural dimensionless time $\tau = \omega t$ leads again to a dimensionless $H_d = \frac{H(t)}{\hbar\omega}$, although the stability thresholds no longer obey the Strutt diagram (see \cite{bender78,mielnik10}).

\newpage
\section{The Motion of Massive Particles: Classification}
\label{sec:optics}

\noindent In quantum optics of {\em coherent photon states}, an important role belongs to the {\em parametric amplification} of Mollow and Glauber \cite{mollow67}. Yet, in the description of massive particles the Heisenberg's evolution of the canonical observables ({\em i.e.}, the `trajectory picture' ) receives less attention, even though it allows to extend the optical concepts \cite{wolf04,wolf12}. This can be of special interest for charged particles in the ion traps, driven by the time dependent fields, coinciding or not with the formula of Paul \cite{paul90}. The most interesting here is the case of quite arbitrary periodic potentials.

For all periodic fields, $\beta(\tau +\footnotesize{\text{T}}) = \beta(\tau)$, the most important matrices \eqref{eq:2} are $u(\footnotesize{\text{T}} +\tau_0, \tau_0)$ describing the repeated evolution incidents. Since they are symplectic, their algebraic structure is fully defined just by one number $\Gamma=\text{Tr}~u(\footnotesize{\text{T}} +\tau_0, \tau_0) $, (without referring to the Ermakov-Milne invariants \cite{ermakov08,milne30,mayo02}). Though the matrices $u(\footnotesize{\text{T}} +\tau_0, \tau_0) $ depend on $\tau_0$, $\Gamma$ does not, permitting to classify the evolution processes generated by $\beta$ in any periodicity interval. The distinction between the three types of behavior is quite elementary:


\begin{enumerate}
\item[{\bf I}]  If $\vert\Gamma\vert < 2$ the repeated $\beta$-periods, no matter the details, produce  an evolution matrix with a pair of eigenvalues $e^{i\sigma}$ and  $e^{-i\sigma}$ ($\sigma \in \mathbb{R}, 0<\sigma<\frac{\pi}{2}$) generating a stable (oscillating) evolution process. It allows the construction of 
the global `creation' and `annihilation' operators $a^+, a^-$ defined by the row eigenvectors of  $u(\footnotesize{\text{T}} +\tau_0, \tau_0)$, but characterizing the evolution in the whole periodicity interval (compare \cite{mielnik11,mielnik10}).

\item[{\bf II}] If $\vert\Gamma\vert = 2$ the process generated by $\beta$ belongs to the {\em stability threshold} with eigenvalues $\pm1$ permitting to approximate a family of interesting dynamical operations ({cf.} the discussions in \cite{mielnik11,mielnik13,mielnik10}).

\item[{\bf III}] If $\vert\Gamma\vert > 2$ then each one-period evolution matrix has now  a pair of real non-vanishing eigenvalues, $\lambda^+ = \frac{1}{\lambda^-}$ with $\lambda^+ = e^\sigma$ and  $\lambda^-= e^{-\sigma}$ producing the squeezing of the corresponding pair of canonical observables $a^\pm$  defined again by the eigenvectors of  $u(\footnotesize{\text{T}} +\tau_0, \tau_0)$ ,  that is, $a^+$ expands at the cost of contracting $a^-$ or vice versa.
\end{enumerate}

The above global data seem more relevant than the description in terms of the `instantaneous' creation and annihilation operators which do not make obvious the  stability/squeezing thresholds. In the particular case of Paul's potentials with  $\beta(\tau ) = \beta_0 + 2\beta_1 \cos \tau$ the map of the squeezing boundary is determined by the Strutt diagram  \cite{bender78}, traditionally limited to describe the ion trapping (in stability areas). Out of them are precisely the {\em squeezing effects} {\bf III}.   To illustrate all this, it is interesting to integrate \eqref{eq:3} for particular case of Paul's potential for $(\beta_0,\beta_1)$ out of the stability domain.

\section{The Mathieu Squeezing}
\label{sec:mathieu}

\begin{figure}[H]
\begin{center}
\includegraphics[width=7cm]{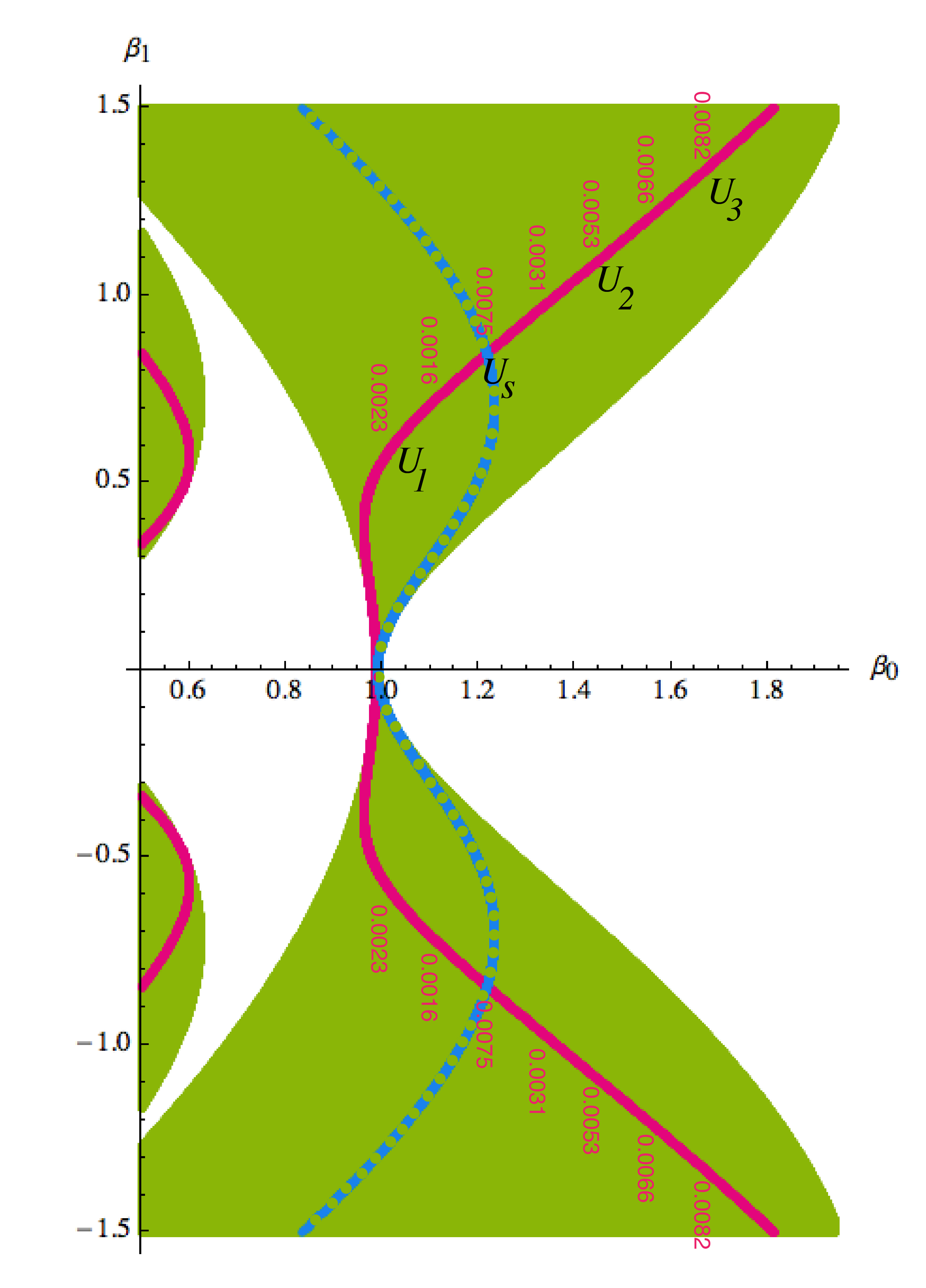}
\caption{\footnotesize The squeezing in Paul's trap. The red line collects the evolution matrices \eqref{eq:2} with $u_{1,2} = 0$ in the second squeezing area on the Strutt map. 
The numbers $\lambda = u_{1,1}$ above, define the {\em squeezing} of $q$ completing the data 
obtained in \cite{mielnik10}. 
The particular matrices $u_1,u_2,u_3$ and $u_s$ obtained for $(\beta_0,\beta_1)$ = \{(1.054,0.646),(1.577,1.231),(1.774,1.454),(1.217,0.844)\}, respectively, are reported in formula \eqref{eq:9}. 
The points on the negative parts of the squeezing trajectory (continuous line), represent the inverted 
{\em squeezing} effects (compare with the reinterpreted Strutt map \cite{mielnik10}). 
\label{fig:1}}
\end{center}
\end{figure}

\noindent As was already found,  the squeezing cannot occur if $\beta(\tau)$ is symmetric in the operation interval \cite{wolf04, wolf12}  we chose to integrate numerically \eqref{eq:3} for Paul's $\beta=\beta_0+2\beta_1\cos\tau$ in $\left[\frac{\pi}{2},\frac{5\pi}{2}\right]$ and $\beta_0,\beta_1$ varying in the second squeezing area of the Strutt diagram (compare \cite{bender78,mielnik10}).

 We then performed the scanning, to localize the evolution matrices 
$u$ yielding the position squeezing. The results shown on  Fig. \ref{fig:1} generalize the numerical data of Ramirez \cite{mielnik10}. 
The continuous (red) line on Fig. 1 represents the $(\beta_0 , \beta_1)$ values for which the evolution matrix $u = u\left(\frac{5\pi}{2}, \frac{\pi}{2}\right)$ has the matrix element $u_{1,2} = 0$, while the interrupted (blue) line contains the $(\beta_0,\beta_1)$ where $u_{2,1} = 0$. 

The intersection of both yields $(\beta_0,\beta_1)$ granting the genuine position-momentum squeezing $q \rightarrow \lambda q$ and $p \rightarrow \frac{1}{\lambda}p$, compare \cite{mielnik10}. 

 Here, the additional data above the red line report slightly different effects when $q$ is squeezed (or amplified) at the cost of distinct canonical variables $a^- = u_{2,1}q + \frac{1}{\lambda} p$. 
As an example,  picked up the four evolution matrices representing various cases of squeezing granted by four pairs $(\beta_0,\beta_1)$ on Fig. \ref{fig:1}. 

\begin{equation}
\label{eq:9}
\begin{split}
u_1 &= \begin{pmatrix} 0.3625 & 0.0023 \\ -1.1147&2.7518 \end{pmatrix}, \quad 
u_2=\begin{pmatrix}0.1757 & 0.0053\\ 3.5018 & 5.7980 \end{pmatrix},\\
u_3 &= \begin{pmatrix}0.2161 & 0.0082\\5.4446 & 4.8334\end{pmatrix}, \quad 
u_s=\begin{pmatrix}0.227570 & 0.007556\\ 0.000447& 4.394266\end{pmatrix}.
\end{split}
\end{equation}

\bigskip
While the matrix $u_s$ at the intersection of both curves in the upper (positive) part 
of the diagram represents the {\em coordinate squeezing} $q \rightarrow \lambda q$, \   $p\rightarrow (1/\lambda) p$ with  $\lambda \approx 0.227$,   the corresponding intersection on the lower (negative) part represents an inverse operation with  $\lambda \approx 4.394$, {\em i.e.} of amplifying $q$ and squeezing $p$.  Henceforth, if the corresponding pulses were successively applied to two pairs of electrodes in a cylindric Paul's trap, then the particle state would suffer the sequence expansions of its $x$ variable with the simultaneous squeezing  of $y$, and then inversely amplifying $y$ but squeezing  $x$, etc. An open question is whether some new techniques of squeezing could appeared by generalizing the operational techniques of high frequency pulses described by  \cite{Itin,Martinez}. 
The reader might feel a bit tired by observing so much effort with such little details.

The boring problem of physical units brings, however, some additional data. For the dimensionless $\tau = \omega t$ the parameter  $\footnotesize{T}$ in \eqref{eq:4} is the period of the oscillating 
 Paul's voltage on the trap wall  $\Phi(t) = \Phi_0 + \Phi_1 \cos \omega t \Rightarrow \beta(\tau) = \beta_0+2\beta_1\cos\tau$. Hence, the same dimensionless matrix $u_s$ of \eqref{eq:9} can be generated  in $\left[\frac{\pi}{2},\frac{5\pi}{2}\right]$, by physical parameters such that:
\begin{equation}
\label{eq:10}
\frac{e\Phi_0}{\omega^2 r_0^2m}=\beta_0, \quad \frac{e\Phi_1}{\omega^2 r_0^2m}=2\beta_1.
\end{equation} 
In case 
of particles with fixed mass and charge, what can vary are the potentials $\Phi$ and the 
physical time $T=\frac{2\pi}{\footnotesize{\omega}}$ of the operations corresponding to 
the dimensionless interval   $\left[\frac{\pi}{2},\frac{5\pi}{2}\right]$. Hence, for any fixed $r_0$, the smaller $\omega$ (and the longer $T$)  the smaller voltages $\Phi_0$ and $\Phi_1$ are sufficient to assure the same result (but only if too weak fields do not permit the particle to escape or to collide with the trap surfaces). For a proton ($m = m_p \simeq \text{1.67} \times 10^{-24}$g) in an unusually ample ion trap of $r_0 = 10$cm and in a moderately oscillating Paul's field with $\omega$ corresponding to a 3km long radio wave, one would have 
$\omega^2 r_0^2 m_p \simeq 10^{-12} \text{g}\frac{\text{cm}^2}{s^2} = \text{1.67} \times 10^{-12}\text{g}\frac{\text{cm}^2}{\text{s}^2} \simeq \text{1.04233}$eV, 
leading to the voltage estimations: 
$\Phi_0 \simeq \text{1.0423}\beta_0 \text{V} \simeq \text{1.268}$V and 
$\Phi_1 \simeq \text{2.0846}\beta_1 \text{V} \simeq \text{1.759}$V. 
In a still wider trap (of an S/F size!) 
$r_0 = 100$cm or, alternatively, for $r_0 =10$cm but the frequency 10 times higher, the voltages needed on the walls should be already 100 times higher. 

While the analytic expressions \cite{frenkel01} could yield more exact results, our computer experiment in fact indicates that the phenomena of  $q,p$--squeezing can happen in the Paul's traps. Yet,  they concern  only the extremely `clean' Paul oscillations, without any laser cooling (crucial for the experimental  trapping techniques  \cite{paul90}), nor any dissipative perturbations.    
Moreover,  the squeezing effects described by matrices \eqref{eq:9} are volatile, materializing itself only in sharply defined time moments, which makes difficult the observation of the phenomenon in the oscillating trap fields.
\bigskip


\section{The Option of Squeezed Fourier}
\label{sec:fourier}

\noindent Mathematically, one of the simplest ways to construct the quantum operations is to apply  sequences of the external  $\delta$-pulses interrupting some continuous evolution process ({\em e.g.} the free evolution, the harmonic oscillation, etc. \cite{ammann97,fernandez92,viola99,viola03}). However,  the method is  obviously limited by the practical  impossibility  of applying the $\delta$-pulses of the external fields. In  case of squeezing, a more regular method could be to compose some evolution incidents which belong to the equilibrium zone ({\bf I}) but their products do not. One of chances is to use  the fragments of time independent oscillator fields \eqref{eq:1} with the elastic forces $\beta = \kappa^2 = \text{const.}$, generating the symplectic rotations: 
\begin{equation}
\label{sprot}
u=\begin{pmatrix}\cos\kappa\tau & \frac{\sin\kappa\tau}{\kappa} \\ -\kappa \sin\kappa\tau & \cos\kappa\tau\end{pmatrix}.
\end{equation}
Their simplest cases obtained for $\cos\kappa\tau=0$ are the {\em squeezed Fourier transformations}
\begin{equation} 
\label{sqF}
u=\begin{pmatrix}0&\pm\frac{1}{\kappa}\\ \mp \kappa&0\end{pmatrix}.
\end{equation}
Following the proposal of Fan and Zaidi \cite{fan88} and Gr\"ubl \cite{grubl89} it is enough to 
apply two such steps with different $\kappa$-values to generate the evolution matrix:
\begin{equation}
\label{squeeze}
u_\lambda = \begin{pmatrix}0&\pm\frac{1}{\kappa_1}\\ \mp\kappa_1&0\end{pmatrix}\begin{pmatrix}0&\pm\frac{1}{\kappa_2}\\ \mp\kappa_2&0\end{pmatrix}=\begin{pmatrix}\lambda&0\\ 0&\frac{1}{\lambda}\end{pmatrix}; \quad \lambda = -\frac{\kappa_2}{\kappa_1}
\end{equation}
which produces the squeezing of the canonical pair: $q \rightarrow \lambda q\ , p \rightarrow\frac{1}{\lambda}p$, with the effective evolution operator: $U_\lambda = \exp[-i\sigma \frac{pq+qp}{2}]; \quad \sigma = \ln\lambda$. It requires, though, two different  $\kappa_1 \neq\kappa_2$ in two different time intervals divided by a sudden potential jump. 
(Here, the times $\tau_1$ and $\tau_2$ can fulfill {\em e.g.} $\kappa_1\tau_1 = \kappa_2\tau_2 = \frac{\pi}{2}$ to assure that both $\kappa_1$ and $\kappa_2$ grant two distinct squeezed Fourier operations in their time intervals.) If one wants to apply two potential steps on the null background, it means at least three jumps ($0 \rightarrow \kappa_1 \rightarrow \kappa_2 \rightarrow 0$). How exactly can one approximate a jump of the elastic potential? Moreover, each $\kappa$-jump implies an energy transfer to the microparticle ({\em cf.} Gr\"ubl \cite{grubl89}). So, could the pair of generalized Fourier operations in \eqref{squeeze} be superposed in a {\em soft way} with an identical end result? In fact, the recent progress in the inverse evolution problem shows the existence of such effects.


\section{Toeplitz Algebra and the Exact Operations}
\label{sec:soft}

\noindent Though the exact expressions \eqref{squeeze} were already known, 
it was not noticed that they can be generated by the 
simple anti-commuting {\em Toeplitz algebra} of $2\times2$ {\em equidiagonal, symplectic matrices} $u$ with $u_{11} = u_{22} = \frac{1}{2} \text{Tr}~u$. It turns out that for any two such matrices $u, v$  their anticommutator $uv+vu$ as well as  the symmetric products $uvu$ and $vuv$ belong to the same family. The  {\em Toeplitz matrices}  inspired a lot of research, {\em e.g.}  \cite{bottcher00,trefethen05,Deift}, though apparently, without paying  attention to their simplest  quantum control  sense. In our case, even without eliminating  jumps in \eqref{squeeze} they give an additional flexibility in constructing the squeezed Fourier operations as the symmetric products of many little symplectic contributions 
\eqref{sprot} with different $\beta$'s acting in different time intervals. Thus, {\em e.g.}, by using the fragments of the symplectic rotations $v_k$ caused by the Hamiltonians \eqref{eq:1} with some fixed $\beta = \beta_k$ in time intervals $\Delta\tau_k ~ (k=0,1,2,\ldots)$, one can define the symmetric product:
\begin{equation}
\label{eq:12}
u=v_n\ldots v_1v_0v_1\ldots v_n
\end{equation}
again {\em symplectic} and {\em equidiagonal} ({\em i.e.} of the simplest Toeplitz class), with $u_{11} = u_{22} = \frac{1}{2}\text{Tr}~u$. Whenever \eqref{eq:12} achieves $\text{Tr}~u = 0$, the matrix $u$ becomes  {\em squeezed Fourier}. The continuous equivalents can be readily obtained. Indeed, it is
enough to assume that the amplitude $\beta(\tau)$ is symmetric around a certain  point $\tau = 0$, {\em i.e.}, $\beta(\tau) = \beta(-\tau)$. By considering then the limits of little jumps $du$ 
caused by applying the contributions $dv = \Lambda(\tau)d\tau$  from the left and right sides, one arrives at the differential equation for $u = u(\tau,-\tau)$ in   the expanding  interval $[-\tau, \tau]$:
\begin{equation}
\label{eq:13}
\frac{\text{d}u}{\text{d}\tau} = \Lambda(\tau) u+u\Lambda(\tau).
\end{equation} 
\bigskip

Its anti-commuting form leads easily to an exact solution. Since $\Lambda(\tau)$ is given by \eqref{eq:3}, equation \eqref{eq:13} becomes
\begin{equation}
\label{eq:14}
\begin{split}
\frac{\text{d}u}{\text{d}\tau} &= \begin{pmatrix}u_{21}-\beta u_{12} & \text{Tr}~u \\
-\beta\text{Tr}~u&u_{21}-\beta u_{12}
\end{pmatrix}\\
& =(u_{21}-\beta u_{12})\mathbb{1}+\text{Tr}~u\begin{pmatrix}0&1\\-\beta & 0\end{pmatrix}
\end{split}
\end{equation}
({\em cf.} \cite{mielnik13,mielnik10}). For the symmetric $\beta(\tau)$, this determines explicitly the matrices $u = u(\tau,-\tau)$ for the expanding $[-\tau,\tau]$ in terms of just one function $\theta(\tau) = u_{12}(\tau,-\tau)$. In fact, since \eqref{eq:14} implies 
the same differential equation for $u_{11}$ and $u_{22}$, {\em i.e.} $\frac{\text{d}u_{11}}{\text{d}\tau}=\frac{\text{d}u_{22}}{\text{d}\tau}=u_{21}-\beta u_{12}$, and since $u_{11} = u_{22} = 1$ at $\tau = 0$, then $u_{11} = u_{22} = \frac{1}{2}\text{Tr}~u = \frac{1}{2}\theta'(\tau)$ for $u$ of any symmetric $[-\tau,\tau]$. Moreover, since $u = u(\tau,-\tau)$ are
symplectic, {\em i.e.} $\text{Det} u = \left[\frac{1}{2}\theta'(\tau)\right]^2-\theta u_{21} = 1$, one obtains

\begin{equation}
\label{u21}
u_{21}=\frac{\left[\frac{1}{2}\theta'(\tau)\right]^2-1}{\theta}.
\end{equation}
\bigskip

Hence, \eqref{eq:14} defines the amplitude $\beta(\tau)$ which had to be applied to create the matrices $u=u(\tau,-\tau)$. Indeed: 
\begin{equation}
\label{eq:15}
\beta u_{12} = u_{21} -\frac{du_{11}}{d\tau}
\end{equation}

and since $u_{12}=\theta$, the $\frac{du_{11}}{d\tau}=\frac{\theta''}{2}$ 
and $u_{21}$ is given by \eqref{u21}, then: 
\bigskip

\begin{equation}
\label{beta} 
\beta=-\frac{\theta''}{2\theta}+\frac{\left[\frac{1}{2}\theta'(\tau)\right]^2-1}{\theta^2}.
\end{equation}
\bigskip

This solves  the  symmetric evolution problem for $u$ and $\beta$ in any interval $[-\tau,\tau]$ in terms of a one, almost arbitrary  function $\theta(\tau)$, restricted by non--trivial 
conditions in single points only. Hence,  \eqref{beta} is  indeed an exact solution of the  inverse 
evolution problem, offering $\beta(\tau)$ in terms of the function $\theta(\tau) = u_{1,2}(\tau,-\tau)$  representing the evolution matrices for the expanding (or shrinking) evolution intervals $[\tau,-\tau]$. 
Note though that the dependence of $u(\tau, \tau_0)$ on $\beta(\tau)$ given by \eqref{beta} in any non-symmetric interval  $[\tau, \tau_0]$ requires still an additional integration of \eqref{eq:3} between $\tau_0$ and $\tau$. Some simple algebraic relations of $\beta$ and $\theta$ are worth attention. 

\bigskip

{\bf  Lemma 2}.  Suppose $\beta(\tau)$ is given by \eqref{beta}, in a certain interval $[-\text{\footnotesize{T}},\text{\footnotesize{T}}]$, where $\theta(\tau)$ is continuous and three times differentiable. The conditions which assure the continuity, differentiability of $\beta$ and the dynamical 
relations between $\theta$ and $\beta$ are then:
\begin{enumerate}
\item[i] At any point $\tau$ where $\theta(\tau) = 0$, there must be $\theta'(\tau) = \pm2$.
\item[ii]  If, moreover, $\theta'''(\tau) = 0$ then also $\beta'(\tau) = 0$.
\item[iii] At any point $\tau$ where $\theta(\tau) \neq 0$ but $\theta'(\tau)=0$, the matrix  \eqref{eq:14} for $[-\tau,\tau]$ represents the squeezed Fourier transformation with $\beta(\tau)$ at the end points  given by  $\beta(\tau)\theta^2=-\frac{1}{2}\theta''\theta -1$.
\end{enumerate} 

{\bf Proof} follows straightforwardly by applying \eqref{eq:15}. In particular, since \eqref{eq:14} and the initial condition grants $u_{11}=u_{22}=\frac{1}{2}\theta'(\tau)$ then, whenever $\theta'=0$, both $u_{11}=u_{22}=0$ implying $u_{12}=b\neq0, u_{21}=-\frac{1}{b}$; which is the general form of the {\em squeezed Fourier} transformation.  Simultaneously, \eqref{beta} simplifies and 
 the value of 
$\beta(\tau)$ fulfills $\beta(\tau)\theta^2=-\frac{1}{2}\theta''\theta -1 \Rightarrow \beta(\tau)b +\frac{\theta''}{2} +\frac{1}{b}=0 $. In particular, if  $\theta''(\tau)b = -2$, then $\beta(\tau) = 0$.

\bigskip

Certain curious {\em quid pro quo} should  be noted. Without entering into  the phase problems \cite{muga10,muga11,guerrero11}  
we used here (and in \cite{mielnik11,mielnik13,mielnik10})  the simplest case of  Toeplitz algebra (see \cite{bottcher00,trefethen05, Deift}) which solves the inverse evolution problem for  $\beta(\tau)=\kappa^2(\tau)$ in terms of $\theta(\tau)$ without any auxiliary invariants . However, its purely comparative sense should be stressed.  For a fixed pair of canonical variables $q,p$ it does not give the causally progressing process of the evolution, but rather compares the evolution incidents in a family of expanding  intervals $[-\tau, \tau]$.  Should one like to follow the causal development 
of the classical/quantum systems, the Ermakov-Milne equation \cite{ermakov08,milne30} might be useful. An interrelation between both methods waits still for an exact description. It is not excluded that the anticommutator algebras can help also in some higher dimensional canonical  problems. \footnote{It seems truly puzzling that this extremely simple case of anti-commuting Toeplitz algebra was never associated with the variable oscillator evolution.} 

\section{The Simplest Cases}
\label{sec:simplest}

\noindent As already checked, there exist polynomial models of $\theta(\tau)$ \cite{mielnik13} making possible the soft generation of the squeezed Fourier transformations (no sudden jumps!). The polynomials of $\tau$, however, are just a formal exercise. As it seems, empirically more natural would be to apply the harmonically oscillating $\theta$ functions. As the most elementary case, let us consider the evolution guided by $\theta$'s  with only four frequencies. In dimensionless variables:
\begin{equation}
\label{eq:16}
\theta(\tau) = a_1 \sin \tau + a_3 \sin{3\tau} + a_5\sin5\tau + a_7 \sin7\tau.
\end{equation}

Note that for $\theta(\tau)$ antisymmetric, the corresponding $\beta(\tau)$ defined by \eqref{eq:15} is symmetric around $\tau=0$. The conditions of our lemma for the $\theta$-function  given by \eqref{eq:16} to generate softly the `squeezed Fourier' with $u_{12}= \pm \kappa =b$ at the ends of the symmetric interval $\left[-\frac{\pi}{2},\frac{\pi}{2}\right]$ become then:
\begin{equation}
\label{eq:17}
\begin{split}
\theta' (0) &= a_1 +3 a_3 + 5a_5 +7a_7= 2,\\
\theta \left(\frac{\pi}{2}\right) &= a_1 - a_3 + a_5 -a_7= b,\\
\theta'' \left(\frac{\pi}{2}\right) &= -a_1 +9 a_3 - 25 a_5 = -\frac{2}{b} -2b\beta_0.
\end{split}
\end{equation}

\begin{figure}[H]
\begin{center}
\includegraphics[width=7.5cm]{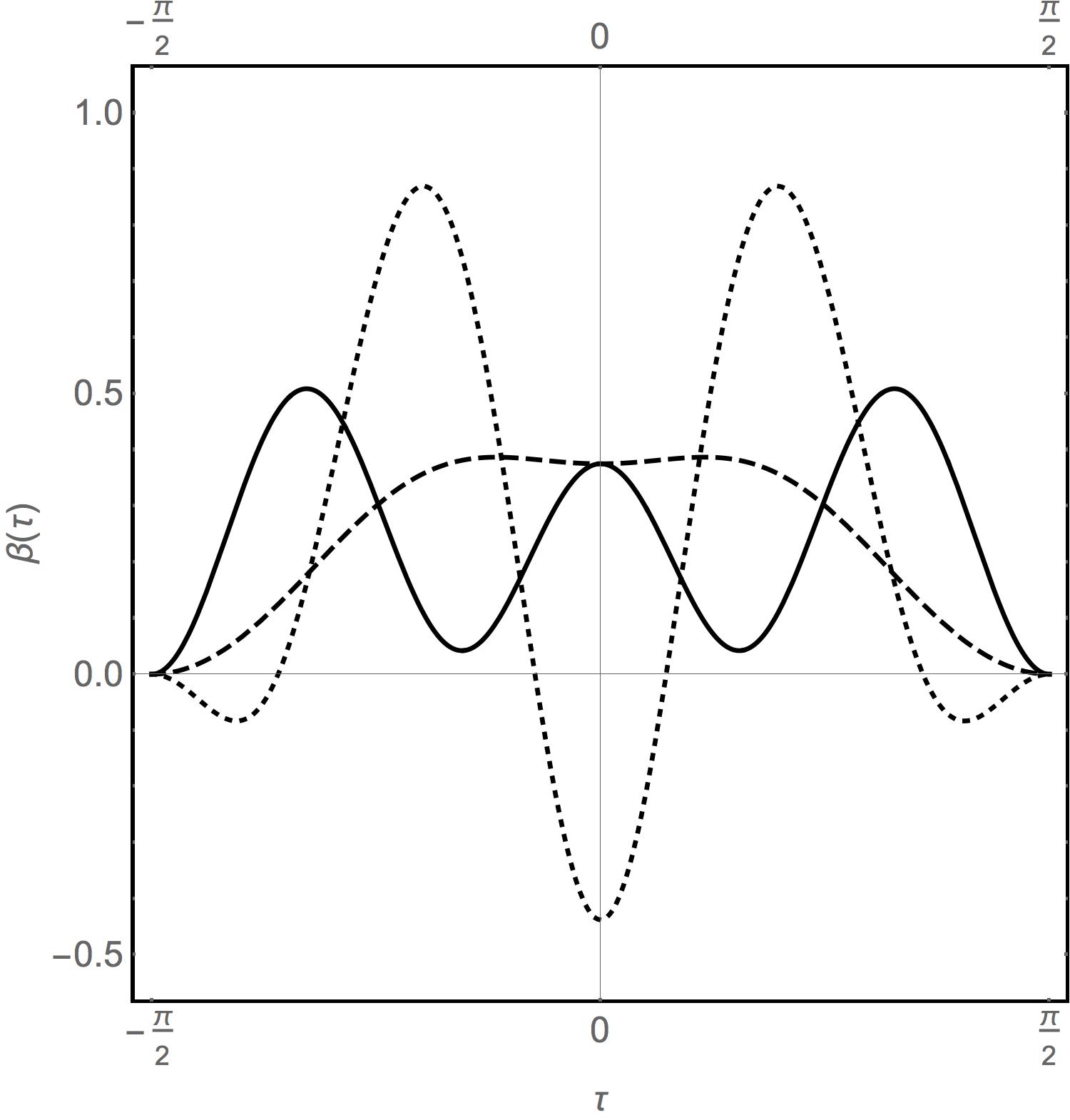}
\caption{\footnotesize  The form of three symmetric $\beta$-amplitudes vanishing softly ($\beta_0=0$) at the borders of the interval $[-\frac{\pi}{2},\frac{\pi}{2}]$ with $b=2,c=-3$ (solid),  $b=\frac{7}{4},c=-3$ (dashed), $b=\frac{9}{5},c=-7/2$ (dotted); all of them assuring the generation of the squeezed Fourier by  the Hamiltonian \eqref{eq:1}. Both solid and dashed curves, which do not cross to the negative values, are adequate to achieve the squeezed Fourier operations by time dependent magnetic fields with $B(t)\sim\sqrt{\beta(t)}$. Each two, superposed softly in two consequtive intervals  $[-\frac{\pi}{2},\frac{\pi}{2}]$ and  $[\frac{\pi}{2},\frac{3\pi}{2}]$ generate the $q, p$ squeezing.} 
\label{fig:2}
\end{center}
\end{figure}

The interrelation between the harmonic $\theta(\tau)$ \eqref{eq:16} and the corresponding physical 
$\beta(\tau)$ given by \eqref{beta} is not completely trivial, but reduces to a purely algebraic problem, 
where the first identity grants the nonsingularity of $\beta$ in 0, the second one defines the magnitude $b$  of the {\em Fourier squeezing} depending on the whole trajectory, and $\beta_0$ defines the symmetric values of the  amplitude $\beta(\tau)$ at $\pm\frac{\pi}{2}$.  Equations \eqref{eq:17} are then fulfilled by:
\begin{equation}
\label{eq:18}
\begin{split}
a_1 = \frac{10-58b-105b^2-bc+10b^2\beta_0}{128b}, \quad a_3 = -\frac{2-74b+35b^2-bc+2b^2\beta_0}{128b}, \\
a_5 =-\frac{-18-22b+21b^2+bc-18b^2\beta_0}{384b}, \quad a_7 = -\frac{6+26b-15b^2+bc+6b^2\beta_0}{384b},
\end{split}
\end{equation}
with $b,c\neq 0$ two suitable real constants. Note that our assumed (antisymmetric) $\theta(\tau)$ represents  $u_{12}(-\tau,\tau)$ for $\tau>0$, while the obtained (symmetric) $\beta(\tau)$ defines the field amplitude in the whole symmetry interval. The examples of the amplitudes with the boundary values $\beta_0$ either vanishing or positive are shown on Fig. \ref{fig:2} and Fig. \ref{fig:4} respectively. 

 We deliberately choose the case of partial $\beta$-amplitudes starting and ending up with $\beta(-\frac{\pi}{2})=\beta(\frac{\pi}{2})=\beta(\frac{3\pi}{2})=0$ to illustrate the flexibility of the method. In fact, we could notice that some programs of frictionless driving seem to exclude the  continuity  at the beginning and  at the end of the transport operation or even assume some sharp steps in the interior. Thus, {\em e.g.},  in  an interesting report Xi Chen {\em et al} 
\cite{muga11} the authors present 
an operation modifying the harmonic oscillator $H_0$ by adding  some perturbation $H_1$, which vanishes before and after the operation, it can also appear or disappear suddenly (likewise in \cite{muga10}). However, the interruption of an adiabatic process by a new potential  which can suddenly  `jump to existence' might be good to achieve the speed and efficiency of {\em frictionless driving} but  not the adiabatic qualities  (see the results of Gr\"ubl \cite{grubl89}).     
\bigskip

\begin{figure}[H]
\begin{center}
$\vcenter{\hbox{\includegraphics[width=7cm]{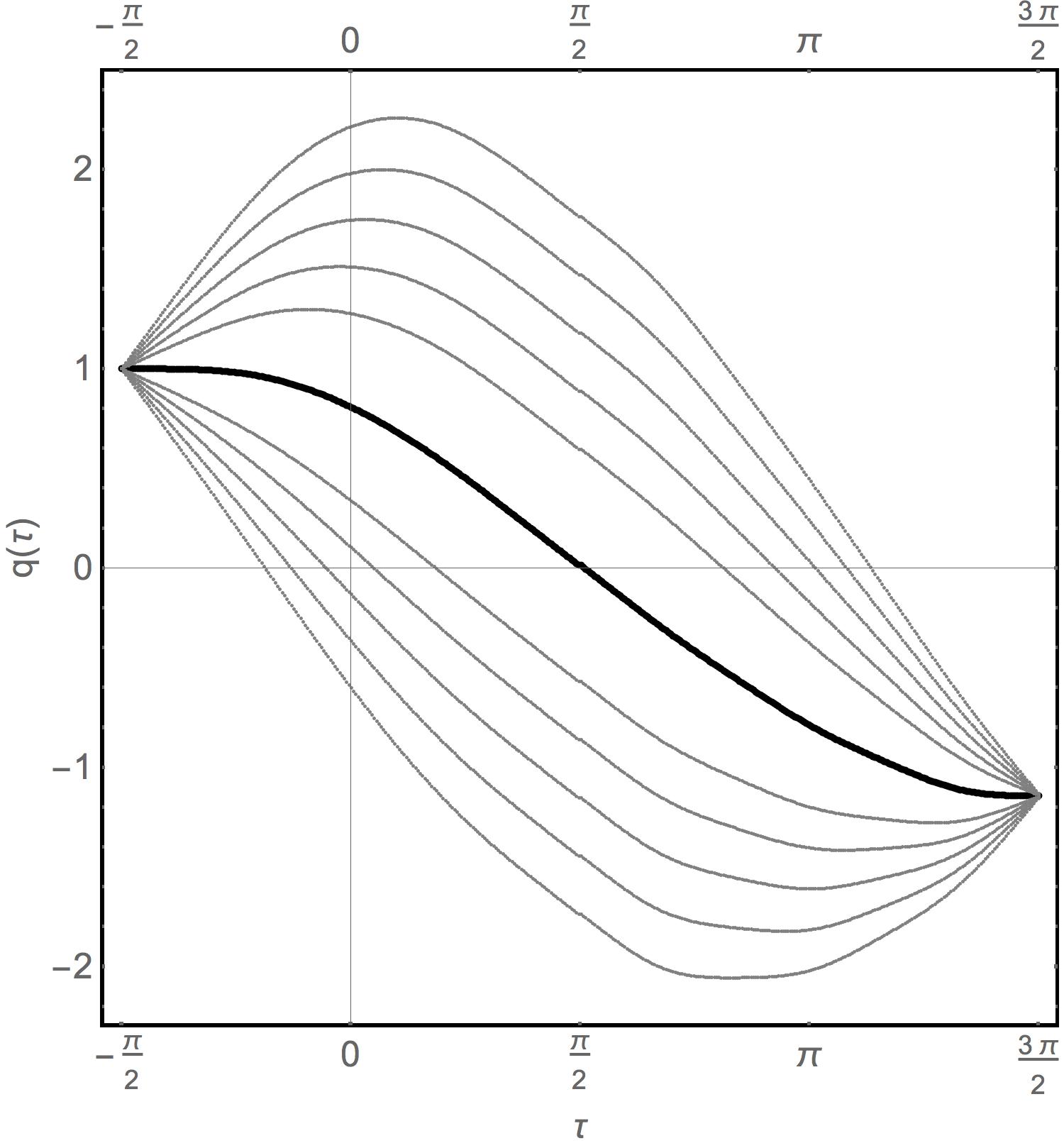}}}$
$\vcenter{\hbox{\includegraphics[width=7cm]{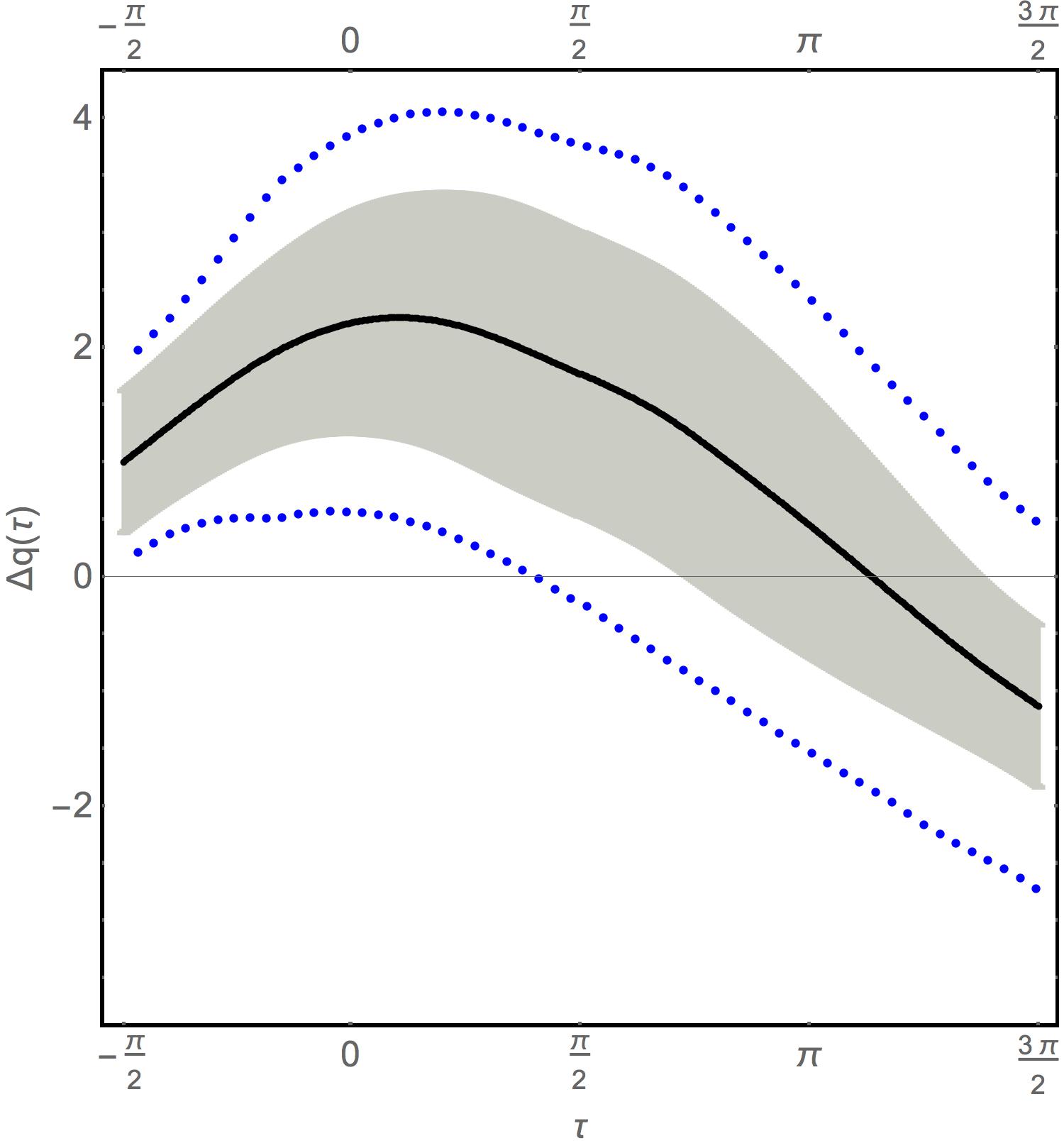}}}$
\\
$\vcenter{\hbox{(a) \hspace{6cm}(b)}}$
\caption{\footnotesize (a) The congruence of dimensionless trajectories  generated by a pair of squeezed Fourier operations induced by two soft $\beta$--amplitudes
of Fig. \ref{fig:2} for $b_{\text I}=2,c_{\text I}=-3$ and $b_{\text {II}}=\frac{7}{4},c_{\text {II}}=-3$ in two subsequent intervals $\left[-\frac{\pi}{2},\frac{\pi}{2}\right]$ and $\left[\frac{\pi}{2},\frac{3\pi}{2}\right]$. The final result is the $q$--amplification with $\lambda =-\text{1.14}$. If generated by a magnetic field in a cylindrical solenoid, it would mean the $\lambda$--expansion of both coordinates $q=x, y$. (b) The position uncertainty $\Delta q$ (the {\em uncertainty shadow}) for the upper trajectory on the  part (a) was determined for the numerically calculated $u_{11}$ and $u_{12}$ according to \eqref{eq:3} with the initial values $\Delta q =\Delta p = \frac{1}{2}$.  So, if the operation is performed for an initial Gaussian packet, in a cylindric Paul's trap or solenoid of dimensionless radius large enough ({\em e.g.} $r_0>10$), this means a little probability of the particle collision with the trap or solenoid wall.
}
\label{fig:3}
\end{center}
\end{figure}

Of course, the $\theta$ and $\beta$-pulses determined by \eqref{eq:16} and \eqref{eq:17}, shown in Fig. \ref{fig:2} are a kind of `extra information' which tells only what operations the time dependent $\beta$-functions  generate at the ends of both  symmetric operation intervals (in this case  the pair of  `squeezed Fourier' in $\left[-\frac{\pi}{2},\frac{\pi}{2}\right]$ and $\left[\frac{\pi}{2},\frac{3\pi}{2}\right]$). This does not yet define the actual trajectory inside both intervals ({\em i.e.} for $\tau\neq -\frac{\pi}{2},\frac{\pi}{2}, \frac{3\pi}{2}$), which must be determined 
by a separate computer simulation. With this aim, we integrated the matrix equation \eqref{eq:3} (not \eqref{eq:13}!)  for $u(\tau,\tau_0)$ with the initial condition $u(\tau_0,\tau_0)=\mathbb{1}$, where $\tau_0=-\frac{\pi}{2}$ starts the first evolution interval, then we continued the integration for the next $\beta$ in the next interval, obtaining a family of $2\times 2$ evolution matrices which draw a congruence of trajectories departing from the beginning of the first and ending up at the end of the second evolution interval  ({\em cf.} Fig.~\ref{fig:3}(a)). As one can see, they indeed paint an image of the pair of `squeezed Fourier' in both subintervals and the coordinate squeezing at the very end.

The above time dependent family  $u(\tau,\tau_0)$  in the sum of both intervals  $\left[-\frac{\pi}{2},\frac{3\pi}{2}\right]$,  permits also to observe the progress of the position and momentum 
uncertainties on the trajectory. As an example, we took one of the the most elementary Gaussian wave functions in $L^2(\mathbb{R})$ centered at $x = q_0$ with the initial velocity $p_0$:
\begin{equation}
\label{eq:19}
\Psi(x,0) = A \exp\left[ i p_0 (x-q_0) \right]\exp\left[-\kappa \frac{(x-q_0)^2}{2} \right] \quad A=\left(\frac{\kappa}{\pi}\right)^{\frac{1}{4}}
\end{equation} For $\kappa=1= q_0=1$, and for varying $p_0$  the packet center 
will draw exactly the family 
of trajectories in Fig.~\ref{fig:3}(a) and the simple calculation with the initial uncertainties 
  $(\Delta q)^2 = (\Delta p)^2 = \frac{1}{2}$, leads to: 
\begin{equation}
\label{eq:22}
|\Delta q(\tau) |^2 = \frac{1}{2}\left[u_{11}^2(\tau) + u_{12}^2(\tau)\right],
\end{equation}

 We then used the square root of \eqref{eq:22} to correct the upper trajectory of Fig. \ref{fig:3}(a) ($p_0=1$) by its {\em uncertainty shadow} (see Fig. \ref{fig:3}(b) ). These results suggests that the main part of the evolving packet is contained within a wider dimensionless belt, {\em e.g.}  $|q (\tau)| < 10$ in the whole evolution interval $\left[-\frac{\pi}{2},\frac{3\pi}{2}\right]$. Characteristically, the uncertainty effects are most visible in the middle of the trajectory, for $\tau=\frac{\pi}{2}$ where two distinct `squeezed Fourier' meet, but they stick to the final `amplified state' at $\tau=\frac{3\pi}{2}$. 
\bigskip

We also checked that the in both cases, i.e. in Fig.~\ref{fig:3} and in Fig.~\ref{fig:5}
our data on $\Delta q$ can provide as well the more detailed statistical information. 
Our initial packet is not an eigenstate of any instantaneous Hamiltonian \eqref{eq:1}, 
but is Gaussian and so are the transformed states $\Psi (x,\tau)$. 
Thanks to the classical-quantum duality, the evolution matrix $u(\tau,\tau_0)$ determines also 
the evolved quantum state  $\Psi$. Some difficulty consists only in expressing $\Psi$ exclusively in 
$x$-representation. Yet, generalizing the already known results \cite{cohen}, we could obtain an explicit 
expression for the $x$-probability density $|\Psi(x,\tau)|^2$ in an arbitrary moment $\tau$:    

\begin{equation}
\label{probability}
|\Psi(x,\tau)|^2=\frac{1}{\sqrt{\pi} \Delta q}e^{-\frac{(x-\langle q\rangle)^2}{|\Delta q(\tau)|^2}}=\frac{\sqrt{2}}{ \sqrt{\pi \left[u_{11}^2 + u_{12}^2\right]}}e^{-\frac{2(x-u_{11})^2}{\left[u_{11}^2 + u_{12}^2\right]}}
\end{equation}
(compare with the formula (17), complement G1 in \cite{cohen}, for the free packet propagation. Our hypothesis is, that our formula \eqref{probability}, not limited to the free packets, is the next step permitting to express 
the probabilities for the Gaussian states in terms of matrices $u_{kl}$  in all 
cases of time dependent elastic forces). 

Our construction differs slightly from the other ones used to generate the squeezing by time dependent oscillator potentials.   
Up to now, the algorithms for the 'soft' squeezing (because of some algebraic difficulties) were designed for non-vanishing initial and final values $\beta=\beta_0>0$. Our development somehow, permits to avoid the difficulty, since it allows $\beta =0$ at the beginning and at the end.  However, the construction deduced from  \eqref{eq:16} allows to generate the same effects for arbitrary initial and final values of $\beta$. As an example, we quote below an analogous case for a non-vanishing pair of initial and final $\beta$ values.  

We therefore checked that the squeezing/amplification of the wave packets can be as  exactly induced by just 
modifying the orthodox harmonic potential to produce the squeezed Fourier operation in  the first operation interval  $[-\frac{\pi}{2},\frac{\pi}{2}]$,    
 conserving the same $\beta=\beta_0>0$ at both ends. The three such modifications are illustrated on Fig. \ref{fig:5}, each of them adequate to join softly the constant time 
independent harmonic oscillator in $\left[\frac{\pi}{2}, \sqrt{ \frac{5}{2}} \pi \right]$ with the same $\beta_0$. We choose the $\beta(\tau)$ represented by the solid line in Fig 4,  (caring to preserve the continuity of $\beta$,  $\beta'$ and $\beta''$ at $\frac{\pi}{2}$). The amplification achieved in the entire  $[-\frac{\pi}{2}, \sqrt{ \frac{5}{2}} \pi ]$ is now much better, $\lambda \simeq -1.71$, the result illustrated on Fig. \ref{fig:5}. 

\begin{figure}[H]
\begin{center}
\includegraphics[width=7cm]{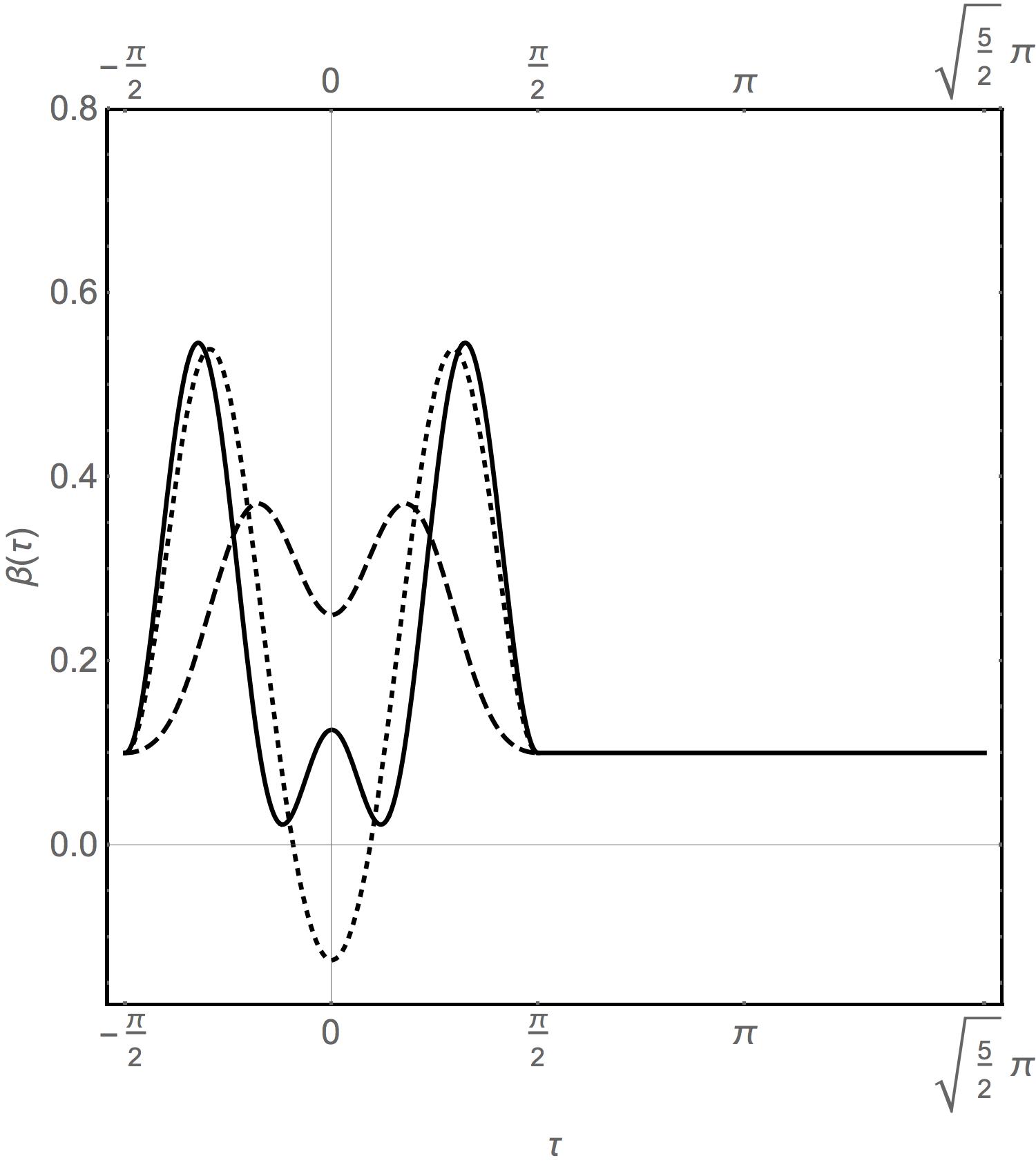}
\caption{\footnotesize The $\beta$-amplitudes satisfying \eqref{eq:17} with the initial $\beta_0=\frac{1}{10}$ and: $b=\frac{43}{20},c=-1$  (solid), $b=\frac{37}{20},c=-2$ (dashed) and $b=\frac{43}{20},c=1$ (dotted). Analogous to Fig.~\ref{fig:3}, both solid and dashed pulses grant the {\em magnetic squeezed Fourier}  in their first action interval $[-\frac{\pi}{2},\frac{\pi}{2}]$. In the next $\tau$--interval $\left[\frac{\pi}{2}, \sqrt{ \frac{5}{2}} \pi \right]$ they all reduce themselves to the constant $\beta_0$ generating the same squeezed Fourier with $b_0=\frac{1}{10}$. Both fragments of 
(I) together produce the amplification with $\lambda\simeq -1.71$.}
\label{fig:4}
\end{center}
\end{figure}

\begin{figure}[H]
\begin{center}
$\vcenter{\hbox{\includegraphics[width=6cm]{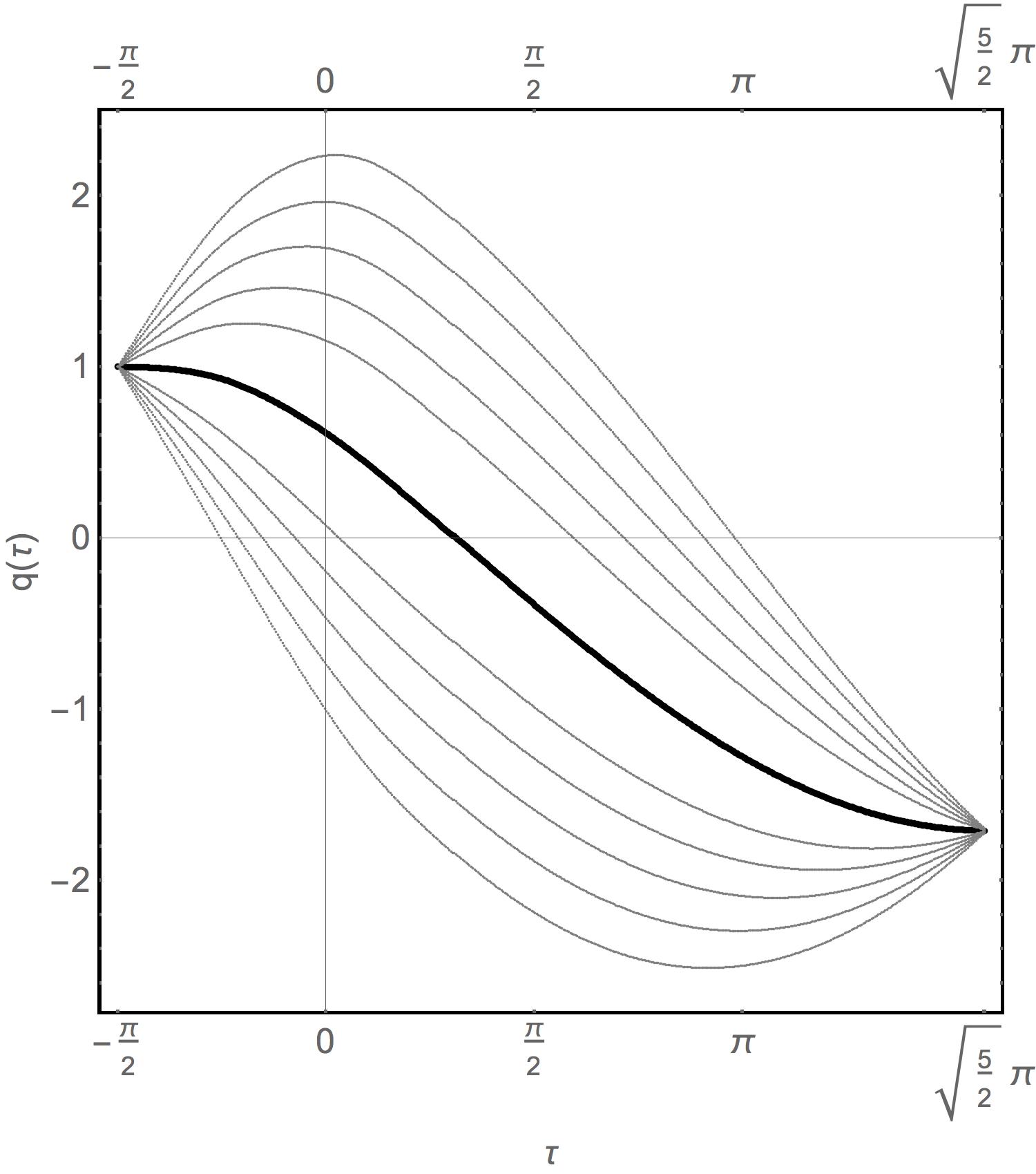}}}$
$\vcenter{\hbox{\includegraphics[width=6cm]{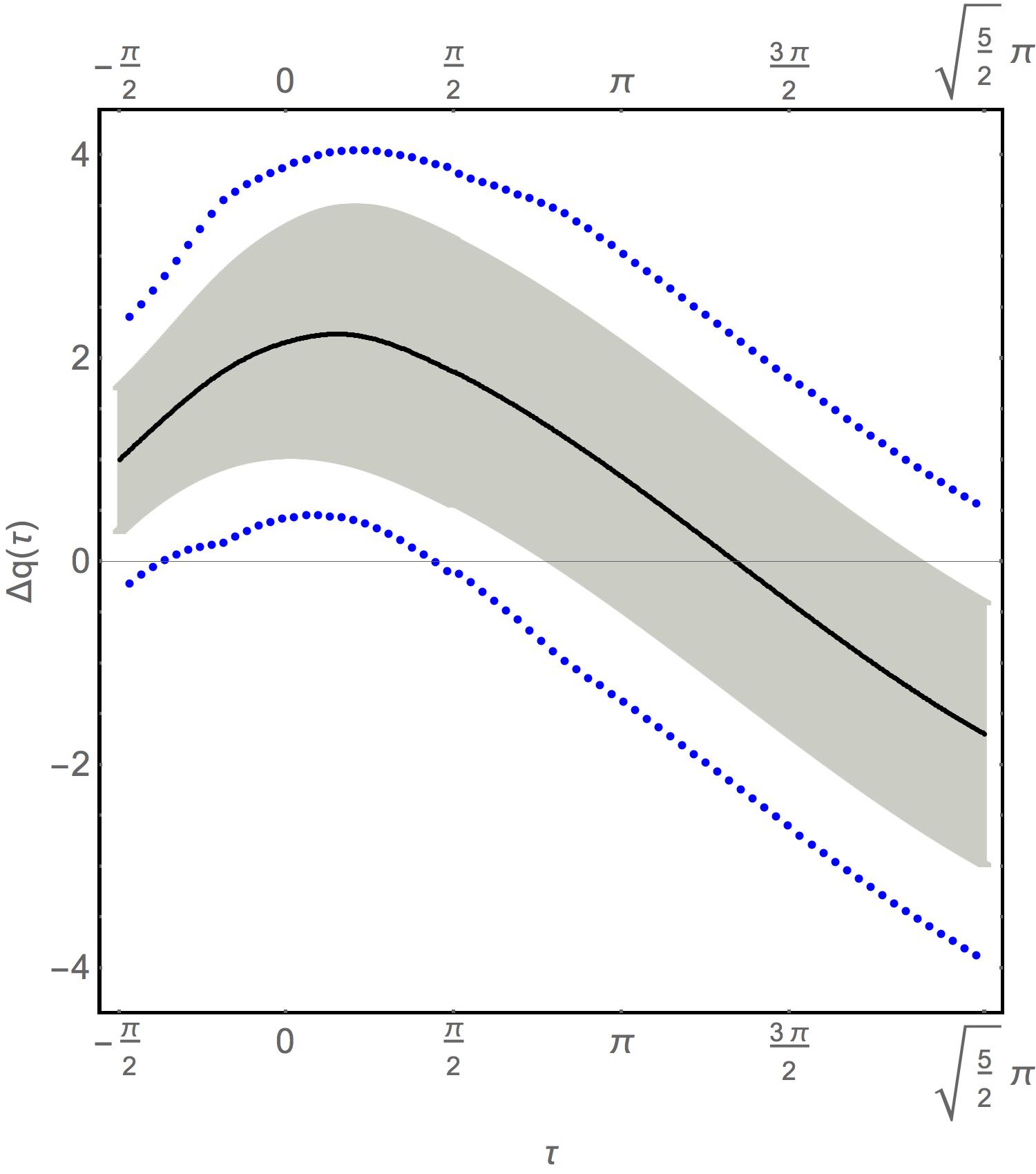}}}$
\\
$\vcenter{\hbox{(a) \hspace{6cm}(b)}}$
\caption{\footnotesize The soft amplification/squeezing with $\lambda \simeq -$1.71 generated by the  dashed amplitude of Fig.~\ref{fig:4}: (a) classical trajectories, (b) the evolving {\em shadow of uncertainty} together with the pair of trajectories marking the 0.999 threshold of packet probability.} 
\label{fig:5}
\end{center}
\end{figure}

\bigskip

All this were just the simplest 1D options of  generating the soft  squeezing of $q$ and $p$. In the axially symmetric Paul's traps this would lead to the simultaneous squeeze (expansion) of both Cartesian coordinates on the 2D plane. The data on Figs. \ref{fig:3}--\ref{fig:5} are applicable also to the magnetic pulses in cylindrical solenoids where, after the separation of the easily integrable rotations, the motion obeys the two dimensional `magnetic oscillators' on the plane orthogonal to the magnetic fields ${\bf B}(t)$. If only the dimensionless amplitude $\beta(\tau)$ of  \eqref{eq:8} coincides with the positive amplitudes of Fig. \ref{fig:2} or Fig. \ref{fig:4}, then the pair of the squeezed Fourier operations will produce an amplification of ${\bf q} = (x,y)$ at the cost of squeezing ${\bf p} = (p_x,p_y)$ (or inversely). 
A certain surprise are the extremely delicate values of the squeezing effects and the corresponding 
electric and magnetic fields  ({\em cf.} Table \ref{tab:1}). Can so weak interactions keep the particle and dictate its unitary transformations? 
Without entering deeper into the  discussion, let us only notice that the extremely weak fields 
could be of importance even in our own existence \cite{penrose94,hagan02,frixione14}.
\bigskip


\section{The Orders of Magnitude}
\label{sec:quantum}
Since the real operation time $\tau T$ can be arbitrarily large, and fields arbitrarily weak,  so borrowing the terminology from \cite{malkin69,malkin73}, we might call it an {\em adiabatic squeezing}. It seems to confirm the existence of the squeezing as purely quantum mechanical phenomenon. 
The fields and times needed in our operations are gathered in Table \ref{tab:1} below.  
\bigskip



Of course, if both squeezed Fourier operations, given by two different matrices $u$ and $v$,  are many times repeated forming
a sequence $uvuvuv\ldots$ then they will cause a sequence of the state amplifications  growing as $\lambda, \lambda^2, \lambda^3,\ldots $ interpolated by increasing sequence of the inverse  (squeezing) effects ~ $\lambda^{-1}, \lambda^{-2}, \lambda^{-3},\ldots$ So, even if the initial effects were very weak ($|\lambda| \simeq 1$), and if the empirical conditions permit,  they could be step by step amplified.  

\begin{table}[H]
\begin{center}\begin{small}
\def\arraystretch{1.8}
\begin{tabular}{l||l|l|l}
\hline
$T$ [sec.]							& 1/1000  					& 1 					& 100 						\\\hline
$q$ [cm] 							& 800$\times10^{-6}$ 		& 25$\times10^{-3}$     & 250$\times10^{-3}$ 		\\\hline
$p$ [g cm $\text{sec.}^{-1}$]		& 1.3$\times10^{-24}$ 		& 42$\times10^{-27}$  	& $4.2\times10^{-27}$		\\\hline
$v$ [cm $\text{sec.}^{-1}$]    		& 800$\times10^{-3}$ 		& 25$\times10^{-3}$ 	& 2.5$\times10^{-3}$ 		\\\hline
$\Phi_{\text{max}}$ [volt] 			& 2.3$\times10^{-6}$ 		& 230$\times10^{-12}$   & 23$\times10^{-15}$		\\\hline  
$B_{\text{max}}$ [G] 				& 15$\times10^{-3}$  		& 154$\times10^{-6}$    & 1.54$\times10^{-6}$		\\\hline
$ F_\text{rad}/F_\text{osc}$ 		& 86.6$\times10^{-27}$     	& 866$\times10^{-30}$    & 8.66$\times10^{-30}$		\\\hline  
\end{tabular}
\end{small}
\end{center}
\caption{\footnotesize The physical conditions for  the amplification/squeezing $\lambda\simeq-1.71$ $\left(\frac{1}{\lambda}\simeq-0.58\right)$,  for a proton in an axially symmetric Paul's trap with $r_0 = 20$cm, or in a cylindrical solenoid,  generated by soft pulses represented in Fig. \ref{fig:4} and Fig. \ref{fig:5}, corresponding to three cases of physical time $T$.  The physical magnitudes of $q, p, v$ in the upper rows correspond to the dimensionless  $q_d=p_d=1$ in definition \eqref{eq:5} for various $T$. The fields are extremely tiny but their amplitudes grow as the operation time becomes shorter.
Orders of magnitude of $\Phi_{\text{max}}$ and  $B_{\text{max}}$ for the subsequent operation times  are quoted in 5th and 6th rows. The last row reports the average ratios of the Abraham-Lorentz radiative force to the time dependent oscillator forces in all operations.} 
\label{tab:1}
\end{table}

\newpage
Of course, all this are merely the preliminary evaluations, valid only when the physical size of traps or solenoids is sufficient. If we are interested in so weak fields, it is not in order to develop a new branch of quantum technology, but rather to check the reality of QM states and operators at the most basic level, of delicate, semi classical and quasi-static operations  unpolluted by radiative phenomena. To observe this, it is of course insufficient to limit attention to the strong but slowly changing or even static fields which can cause the quantum jumps and radiative cascades.  
\bigskip

 The question thus open, for strong or weak fields, is the 
radiative pollution caused by the driven object itself. Even if the field pulses are periodic, but frequency very low,  the 
formalism of Floquet Hamiltonians does not seem adequate to estimate the instantaneous effects.  In spite of all doubts (the runaway solutions etc.), we decided to check the possible magnitude  
of the Abraham--Lorentz radiative force (which seems quite natural in the `trajectory doctrine'). 
While the ordinary force of the variable oscillator trajectory is simply ${\bf F}_\text{osc}=m{\bf  \ddot{x}}$,
the hypothetical radiative force is expressed as ${\bf F}_\text{rad} = m\sigma {\bf \dddot{x}}$ , where $\sigma$ is the particle dependent 'characteristic time' \cite{jackson99}. Using now the definitions of 
dimensional quantities \eqref{eq:5} we can compare the magnitudes of the conventional and 
radiative forces for the squeezing operations in our table, finding the radiative ones extremely small but slowly increasing for shrinking  $T$  ({\em i.e.} higher frequencies) on our scale ({\em cf.} the lowest line in Table \ref{tab:1}).  Further  problems in the `trajectory approach'  are still open. 
\bigskip

We noticed that  in works using the Ermakov--Milne invariants 
the operations are supposed to be faster but still `frictionless' \cite{muga10,muga11,guerrero11}.  In this aspect, our contribution is slightly different. Its main point is the use of te Toeplitz algebra to obtain the exact solutions.  The other trends use the suggestive idea of {\em  frictionless driving},   which transports the states without 
changing the eigenvalues of certain invariants. But could it be adiabatic enough to reduce the radiative pollution? Could the more general variational methods, following {\em e.g.}  \cite{carlini06,berry09,muga11} be applied at the level of $\theta$-function?  The question whether Ermakov can help Toeplitz or {\em vice versa} is still open. Moreover, the control problems by time dependent fields are certainly not limited to the microscopic scale.\footnote{A non trivial  macroscopic analogue would be a skillful waiter running with a plate full of liquid without spilling a drop.}. 
Some other practical questions can neither be dismissed.  In fact, in most papers on the state evolution the results are presented in rather abstract form. Yet, the experimental works are seldom worried  
by abstract aspects. Hence, while offering certain exact elements, our contribution  is still 
at some distance from the realistic laboratory techniques. 
\bigskip

\section{Imperfections and Open Problems}
\label{sec:imperfections}\noindent\hspace{0.7cm} 

 {\em The troubles with geometry}. In many  laboratories the  
techniques used to keep and cool the ions are adequate to  study the atomic structures but 
insufficient for wider purposes, since the time dependent oscillator potential is created only in 
a strictly local scale {\em e.g.} in an immediate vicinity of the central axis of a quadrupole trap, 
formed in some cases just by four metal bars \cite{senko02}.\footnote{In the recently proposed {\em charged resonator} \cite{resonator13} the pulsating oscillator field is approximated only in small vicinity 
of a single point.} The technical chance to approach the oscillator fields in a wider space to control the  unitary evolution occurs either in the conventional  or in cylindrical Paul's traps with perfectly 
hyperbolic surfaces or else, in the interiors of the cylindrical solenoids, in both, if the operation                                                              
area is wide enough. If so, however, then the controlling  field propagation in the trap interior 
becomes an additional problem.  
\bigskip

{\em The relativistic corrections}. The propagation of the electromagnetic signals  in the Paul's traps  on the trap surfaces or interiors is usually disregarded due to the very small trap size. In reality, however, even some very slow potential changes on the trap surfaces must produce the field corrections  starting from  $\frac{1}{c}$ ({\em post Newtonian}) terms in the EIH approximation \cite{infeld60}. The chances to create the quasi static $\simeq \frac{1}{c^2}$ ({\em post-post Newtonian}) conditions were considered in \cite{mielnik13}. Below, we shall face the similar problem for softly changing  magnetic fields. The  time dependent, homogeneous magnetic field $B(t)$ in the cylindric  solenoid does not fulfill the Maxwell equations. Yet, it obeys  the sequence of EIH approximations. To evaluate the errors, let us  look for the exact time dependent vector potentials of the cylindrical solenoid in the form:
\begin{equation}
\label{eq:23}
{\bf A}({\bf x},t) = \frac{1}{2}B(r,t){\bf n}\times{\bf x} = \frac{1}{2}B(r,t)\begin{pmatrix}-y\\x \end{pmatrix},
\end{equation}
where the magnetic field $B$ instead of depending only on $t$, could also depend on radius $r$ on the perpendicular solenoid section. To assure the relativistic sense of \eqref{eq:23} we must assume $\square {\bf A} = \frac{4\pi}{c} {\bf j} = 0$ where $\square= \frac{1}{c^2}\frac{\partial^2}{\partial t^2}-\Delta$. As easily seen, the application of the Laplacian to the right hand side of \eqref{eq:23} is equivalent to apply the operator $ D \equiv \frac{\partial^2}{\partial r^2}+\frac{3}{r}\frac{\partial}{\partial r}$ to  $B (r, t )$ alone. Hence, the vanishing of the d'Alambertian $\square {\bf A}$ means:
\begin{equation}
\label{eq:24}
\left[\frac{1}{c^2}\frac{\partial^2}{\partial t^2}-D\right]B(r,t) = 0.
\end{equation}

To assure the analytic shape of $B(r,t)$ around the $z$-axis we look for the solution in the form:
\begin{equation}
\label{eq:25}
B(r,t) = B_0(t)+B_2(t)r^2+B_4(t)r^4+\cdots
\end{equation}
where $B_0(t) =B(t) $ is the homogeneous quasistatic approximation. Since $Dr^{2n}=4n(n+1)r^{2(n-1)}$, \eqref{eq:24} and \eqref{eq:25} after short calculation yield:
\begin{equation}
\label{eq:26}
B_{2n}(t) = \frac{1}{4^n n! (n+1)!}\frac{1}{c^{2n}}\frac{\partial^{2n}}{\partial t^{2n}}B(t).
\end{equation}

By introducing now a dimensionless time $\tau = \frac{t}{T}$, where $T$ is some conventional time unit corresponding to the laboratory observation, and by writing \eqref{eq:26} in terms of the time derivatives $\frac{\partial}{\partial\tau}$ one can reduce it to:
\begin{equation}
\label{eq:27}
B(r,t) = B(t) + \frac{1}{8}\left(\frac{r}{cT}\right)^2\frac{\partial^2 B(t)}{\partial^2 \tau} + \frac{1}{24}\left(\frac{r}{cT}\right)^4\frac{\partial^4 B(t)}{\partial^4 \tau}+\cdots
\end{equation}
We kept here the $B(t)$ depending on the real time $t$, in order to assure that all derivatives $\frac{\partial^n}{\partial\tau^n}B(t)$ will be expressed in magnetic field units. The curious property of this formula is the absence of terms $\sim \frac{1}{c}$; the field propagation law \eqref{eq:24} is solved exclusively in terms of extremely small contributions $\sim \frac{1}{c^2}$.  It suggests that the superpositions of delicate wave fronts running towards the solenoid center create a good approximation of the quasistatic theory. 

\bigskip
\noindent\hspace{0.7cm} {\em The control of the currents}. Even so, this does not explain how to create (or at least approximate) the first magnetic step $B(t)$,  to {\em wake up} the whole iterative series \eqref{eq:27}.  (That is, how to induce  the homogeneous surface currents 
 which do not depend on $z$, but depend on time in any desired way). 
In the static case, the magnetic field $B$ in the solenoid is generated by the stationary current ${\bf j}$ circulating around the solenoid surface. The well known result obtained by integrating the magnetic field along the closed contour surrounding the solenoid wires tells that $\bf B$ is defined by total current circulating per unit of the $z$-axis, {\em i.e.}, 
$B =  \frac{4\pi}{c}\frac{\Delta I}{\Delta z}$ . 
However, how to produce the circulating currents depending on time, but homogeneous on all surface sections {\em i.e.} independent of $z$?  If the solenoid was constructed as a single spiral wire around the cylindrical surface, connected at the both extremes to the potential difference $\Phi(t)$, then even a soft change of $\Phi$ will propagate along the solenoid as a pulse of the current, creating inside  the soft but $z$-dependent fields, instead of the quasistatic $B(t)$. Perhaps, a good approximation could be --instead of a single wire-- to cover the cylindric surface by some number of shorter wires connected with a common source of variable voltage? Perhaps, it is not the only  solution.

\bigskip
\noindent\hspace{0.7cm}{\em The model of rotating cylinder}. A different  idea is inspired by an  example described by Griffiths \cite{griffiths99}. The cylindrical surface of non-conducting material ({\em e.g}. glass), of radius $R$ is charged uniformly by surface density $\sigma$, so that each circular belt of 1cm height contains the charge of $R\sigma$. The experimental challenge 
is not extraordinary.  If the cylinder has the radius $R = 20$cm and rotates with frequency $\omega = 1\text{s}^{-1}$ around its axis $z$ and has the charge 1C on each 1cm horizontal belt, then it will produce inside the homogeneous magnetic field ${\bf n} B$ of intensity:
\begin{equation}
\label{eq:28}
B = \frac{4\pi}{c}\omega R\sigma\simeq 1.25\text{G}.
\end{equation}
at least in the post-post-Newtonian approximation.  Hence, by employing the softly changing angular velocity $\omega=\omega(t)$ one can generate the practically homogeneous magnetic field ${\bf n}B(t)$ of the quasistatic environment described by \eqref{eq:27}. Will such techniques work? 

\bigskip
\noindent\hspace{0.7cm}{\em The time control}. In order to perform an operation induced by variable fields on a quantum state, the microobject must be submitted to the action of these fields in the exact time interval between the operation beginning and the operation end. In case of  operations induced by the time dependent magnetic fields, it means that the charged particle must be injected in the known initial state to the solenoid at the exact beginning of  the squeezed Fourier operation  represented on Fig. \ref{fig:2} and the result of the operation must be checked again in the well defined moment, after one or  several applications of the field pattern. The need of this double time  synchronization is almost never considered in papers on quantum control\footnote{We acknowledge the discussions with Dr. Jan Gutt from Phys. Dept. of Polish Acad. Sci.}.  

What one can imagine is a long but finite solenoid; then the particle injected at a
precisely controlled moment, at a point $z=z_0$ in one solenoid end, with a certain velocity $v_z$. Now, the particle wave packet propagates, changing its shape on the subsequent perpendicular solenoid sections, until arriving to the other end, during the time corresponding exactly to one or more squeezing operations. Once arriving there, it is received by a measuring device ({\em e.g.} a photographic plate) registering its position on the new orthogonal plane. Certain errors of this scheme are inevitable. In fact, if the registering screen has a granular structure, then the final particle position will be known only with accuracy to the distance between the mesoscopic detectors. Moreover, since for a particle injected at a given $z_0$ (solenoid beginning), the velocity of flight along the $z$-axis must obey the uncertainty relation $[z,p_z]=i\hbar$ so, the time of flight might fail to reproduce exactly the time of the squeezing operation. 
An interesting aspect is, however, that if the squeezing between the initial and final particle state on two orthogonal planes consists in coordinate amplification, {\em i.e.}, $\tilde{x} = \lambda x_0$ and $\tilde{y} = \lambda y_0$ and the momenta shrinking,
$\tilde{p}_x=\frac{1}{\lambda}p_x$ and $ \tilde{p}_y=\frac{1}{\lambda}p_y$ (with $\vert\lambda\vert>1$), then the situation is almost equivalent to the non-demolishing measurement \cite{nondemolishing78,nondemolishing80}, but with an important numerical difference, that even from an imprecise measurement of the final coordinates $\tilde{x},\tilde{y}$, the data about the initial ones $x_0, y_0$ will be recovered with much smaller errors $(\Delta x_0, \Delta y_0)= \frac{1}{\vert\lambda\vert}\left(\Delta\tilde{x},\Delta\tilde{y} \right)$.


\bigskip
\noindent\hspace{0.7cm}{\em The neglected perturbations}. In all our calculations we considered the pure particle states, evolving in a slowly changing external fields, without taking corrections for  traces of matter in the ion traps or in the solenoids. So, how perfect must be the trap vacuum, to realize indeed our squeezing operations? Moreover, we have neglected the possible direct packet reflection from or absorption by the walls of the laboratory. So, how large must be the ion trap or the solenoid to make any of these troubles insignificant? Moreover,  in case of possible particle absorption by a surface, the problem indeed leads to the fundamental question of the `time operator' which persists without a truly convincing solution even in case of flat surfaces. We can hope only that for the traps wide enough, our {\em soft operations} bring something of interest to the quantum control theory. We also did not consider the variety of hypothetical modifications described by the dissipative mechanisms of Lindblad,  Gisin and Percival \cite{lindblad76,gisin92,gisin99}. Be all this open problems, some other controversial aspects remain to be discussed.


\section{The Fundamental Aspects in Little}
\label{sec:little}

\noindent In spite of imperfections, we feel attracted into a kind of {\em what if story}. The problem is, whether the existing difficulties to achieve the squeezing  are purely technical or they mean some fundamental barrier? If no barrier exists, and the operations can be indeed performed (or at least approximated), the implications could be of some deeper interest.

If the squeezing of the wave packets in $L^2(\mathbb{R}^n)$ defined by $\psi(x) = (\sqrt{\lambda})^{-n} \psi\left(\frac{x}{\lambda}\right)$ (in the lowest dimensions $n = 1, 2, 3)$ with $\vert\lambda\vert <1$ could be achieved as a unitary evolution operation, it would imply that no fundamental limits exist to the possibility of shrinking the particle in an arbitrarily small interval (surface, or volume). Some authors believe that such localization must fail at extremely small scale `below Planck distance'. It is indeed difficult to dismiss {\em a priori} the doubts. However, in many fundamental discussions, the {\em Planck distance} is used as a magic spell which permits one to formulate almost any hypothesis free of  consistency conditions. Yet, if one truly believes that QM is a linear theory, then even the most concentrated wave packets are just the linear combinations of the extended ones. So, the hypothesis about the new microparticle physics, below some exceptional limits, can hardly be defended without modifying everything. In particular, the theories of {\em non commutative geometry} in which the space coordinates fulfill $[x,y]=\sigma\neq0$ as a fundamental identity, could not be constructed. It is enough to note that then the simultaneous squeezing transformation $x\rightarrow\nu x$ and $y\rightarrow\nu y$ where $\nu\neq0$ would ruin the non-commutative law.

As essential consequences would follow from the inverse operations in which the initial wave packet could be amplified. Some time ago, a group of authors studied the properties of the radiation emitted from the {\em extended sources}, asking whether some properties of such sources can be reconstructed from the emitted radiation \cite{rzazewski92}. However, each extended  packet is a linear combination (superposition) of the localized ones. The question then arises, what would happen if some experiments could {\em fish} in the emitted photon state some components corresponding to the localized parts of the initial source? Would the initial state  be `reduced' to one of its localized emission points (as in the {\em delayed choice experiment} of J.A. Wheeler? The idea seemed unreal, so the authors of \cite{rzazewski92} worried rather about the momenta than space localization of the extended source. (The situation, however, could be different in case of  the amplified  particle states on a plane orthogonal to the symmetry axis of some solenoid or  ion trap.) 

In fact, the possible state amplification in 2D, (e.g. around the  solenoid  axis),  would lead to unsolved  reduction problems  no less  
challenging {\em interaction free measurement} of Elitzur and Vaidman \cite{elitzur93}.  Here, the finally measured position ${\bf \tilde{q}} = \lambda {\bf q}$ would be an observable commuting with the initial one.  Hence, the observer performing an (imperfect) position measurement in the future could obtain {a posteriori} (in a cheap way) the more exact data about its position in the past. The problem is, whether it repeats the scenario the `non-demolition' measurement \cite{nondemolishing78,nondemolishing80}? If so, does the reduction of the wave packet affects also the particle state in the past, in a new case of  `delayed choice' ?  However, the localization must cost some energy ({\em cf.} Wigner, Yanase {\em et al} \cite{yanase64,jauch67,yanase71}). One might suppose that the energy needed to localize the amplified packet in the future was provided by the reserves in the screen grains (or whatever medium in which the particle was finally absorbed). Yet, this would require an assumption that the relatively low energy invested  in the future should be pumped into some higher energy needed for  more precise localization  in the past. This  seems impossible --not just due to the causality paradox, but also,  due to the energy deficit!

Some fundamental aspects  of quantum mechanics seem in opposition. An undeserved return to the old discussions?  Or perhaps, good news for {\em quantum tomography} (because it might mean the possibility of arbitrarily precise scanning of the future wave packets without an excessive energy pumped into the past). Yet,  one should not forget that all our techniques were based on exploring the evolution matrices \eqref{eq:2} which obey a strictly linear, orthodox QM. Is this theory indeed true? Whatever the answer, it looks like the low energy phenomena might be as close (if not closer) to the fundamental problems of quantum theory as well as high energy physics. 

\bigskip
The authors are indebted to their colleagues at Physics Department in CINVESTAV Mexico, at the summer school in Bialowie$\dot{\text{z}}$a, Poland and at Quantum Fest Conference of UPIITA-IPN Mexico for their interest and helpful remarks. The support from the CONACYT project 152574, is acknowledged.


\bibliographystyle{unsrt}
\bibliography{references}

\end{document}